# Observation of above-threshold dissociation of $Na_2^+$ in intense laser fields

Qun Zhang[1*], John W. Hepburn[2], and Moshe Shapiro[2]

[1]*Hefei National Laboratory for Physical Sciences at the Microscale (HFNL) and Department of Chemical Physics, University of Science and Technology of China, Hefei, Anhui 230026, P. R. China*

[2]*Departments of Chemistry and Physics & Astronomy, and The Laboratory for Advanced Spectroscopy and Imaging Research (LASIR), The University of British Columbia, Vancouver, B.C., V6T 1Z1, Canada*

*Author to whom all correspondence should be addressed: qunzh@ustc.edu.cn

We report the observation of above-threshold dissociation of a multielectron molecular system, $Na_2^+$. The process is due to continuum-continuum transitions in a field-dressed molecular continuum. Three well resolved fragmentation channels with different kinetic energy release patterns have been detected, of which, two are associated with ordinary photodissociation via "light-induced potentials" into Na(3s) + $Na^+$ and Na(3p) + $Na^+$, and one with continuum-continuum transitions and above-threshold dissociation into Na(3p) + $Na^+$. We show that when we dress this molecule with a field whose intensity is ~3 × $10^{12}$ W/cm$^2$, above-threshold dissociation prevails over other processes, such as field-ionization followed by Coulomb explosion.

When subjected to intense laser fields above a certain peak power ($>10^{11}$ W/cm$^2$), small molecules are expected to exhibit a wealth of new phenomena, such as above-threshold dissociation (ATD), molecular bond softening, vibrational population trapping, and high-order harmonic generation [1-9]. These interesting effects, which are a result of the nonlinear response of molecules to strong fields, arise from multiphoton absorption or emission processes.

ATD is a "half-collision" process in which molecules absorb more photons than necessary to dissociate [10]. The experimental signature of ATD is the appearance of a series of peaks, evenly spaced by one quantum of photon energy, in the kinetic-energy distribution of the photofragments [11]. To date, ATD has only been observed in the simplest single-electron system – that of $H_2^+$ ($D_2^+$) [12]. In this system the electronic states are energetically well separated and the analysis of the ATD peaks is a straightforward one, involving only the two lowest-lying potential curves. For heavier molecules, the presence of many electrons makes it much more difficult to unambiguously determine whether certain structures observed in the kinetic energy distribution are indeed being due to ATD [13].

The situation is clearer in the $Na_2^+$ where a recent theoretical investigation of the photodissociation in intense laser fields using a four-state model clearly pointed out to the existence of three dissociation channels, of which one can only arise from ATD [6]. A follow-up experiment [7] verified two such channels, while the expected ATD channel did not materialize. Rather, two other channels were observed, the origin of which was interpreted as field-ionization followed by Coulomb explosion [7].

In this Rapid Communication, we report the observation of the "missing" ATD process in the photodissociation of $Na_2^+$. We report this as the case where ATD was observed in a multielectron molecule.

The photodissociation of $Na_2^+$ by an intense laser field can lead to the formation of a $Na^+$ ion and a Na atom, or to be accompanied by ionization to yield two $Na^+$ ions which separate due to "Coulomb explosion". In order to observe the desirable ATD effect, we have chosen the laser field intensity to be $\sim 3 \times 10^{12}$ W/cm$^2$. This choice of field intensity is based on two considerations: (1) The exploratory calculations [6] have shown that ionization begins to compete with dissociation only at field intensities of $5 \times 10^{12}$ W/cm$^2$ and above, while the threshold for the predicted ATD channel occurs at $3 \times 10^{12}$ W/cm$^2$. (2) Previous observations [7] using a *bound-continuum* coupling scheme demonstrated that above $3 \times 10^{12}$ W/cm$^2$ the field ionization followed by Coulomb explosion is the dominant channel, thereby suppressing the ATD process.

As in other strong-field photodissociation studies, such as that of $H_2^+$ (though exceptions do exist – see, e.g., Ref. [14]), we prepare the $Na_2^+$ molecular ion via a multiphoton ionization (MPI) process starting from the electronic ground state of the neutral $Na_2$ precursor, using the same laser pulse that subsequently fragments the ion. Although this single pulse excitation scheme may sometimes lead to ambiguities in distinguishing between the ionization of the neutral molecule from the dissociation dynamics of the ionic molecule [3,15], as explained below, the present experiment does not suffer from such ambiguities.

The experiment was conducted in a molecular beam environment in which vibrationally cold $Na_2$ molecules are formed during the expansion of pure sodium vapor from an oven heated to 900 K. The cold $Na_2$ molecules are intersected perpendicularly with a mode-locked-amplified Ti:Sapphire femtosecond laser (150 fs FWHM, 48 $\mu$J/pulse at 800 nm) in the interaction region of a time-of-flight (TOF) mass spectrometer. Ions produced by the MPI and/or the fragmentation processes are mass-selectively detected by a microchannel plate. A 500-mm achromatic lens is used to focus the laser to a spot of ~100-$\mu$m-dia., which translates to a peak intensity of ~$3 \times 10^{12}$ W/cm$^2$. A half-wave plate is employed to set the laser polarization to be parallel to the TOF axis. The spectrometer was calibrated by fitting the measured arrival times of the $Na^+$, $Na_2^+$, and $Na_3^+$ ions to derive a reference kinetic energy for the $Na^+$ fragments of interest.

Fig. 1 displays the observed TOF mass spectrum of the $Na^+$ fragments. The predominant central peak is attributed to the MPI of sodium atoms contained in the molecular beam. The remaining three pairs of peaks, distributed evenly about the central peak, are due to photofragmentation of the $Na_2^+$ molecules along the TOF axis. Such a fragment distribution is associated with a $\Sigma - \Sigma$ type transition in the parent $Na_2^+$ molecule [10]. Similar observations have been described for $H_2^+$ [10,12,16] and $Na_2^+$ [7]. The nuclear motion, confined by the strong laser to be directed along the polarization axis, results in two distinct arrival times at the detector. These times correspond to $Na^+$ fragments with initial velocities moving towards or away from the detector. Using the calibration of the TOF mass spectrometer, the total released kinetic energies of the three well resolved photofragmentation channels were determined to be 0.2 ± 0.05 eV (*A*), 0.8 ± 0.2 eV (*B*), and 1.8 ± 0.3 eV (*C*).

To clarify the physics behind the three observed fragmentation channels, we start with a survey of various possible transitions in the *field-free* potential curves of $Na_2^+$. Fig. 2 shows such potential curves computed using a model-potential method [17] for the first eight $\Sigma$ states and the first four $\Pi$ states of $Na_2^+$.

The adiabatic ionization potential of $Na_2$ is ~5.14 eV, requiring absorption of at least four 800-nm photons for ionization. The $Na_2^+$ molecules are prepared in the $1^2\Sigma_g^+$ ground ionic state continuum (indicated by the lowest horizontal dashed-dotted line in Fig. 2) via a direct *non-resonant* four-photon ionization from the ground state of the neutral $Na_2$ molecules. This continuum state serves as the starting point in this study, from which one-photon absorption ($\lambda$ = 800 nm ~ 1.55 eV) can barely lead to dissociation in the continuum of the $1^2\Sigma_u^+$ state. One more photon absorption allows dissociation in the continuum of the $1^2\Sigma_g^+$, $2^2\Sigma_g^+$, $3^2\Sigma_g^+$, and $1^2\Pi_g$ states, while the three-photon absorption opens up the $1^2\Sigma_u^+$, $2^2\Sigma_u^+$, $3^2\Sigma_u^+$, $4^2\Sigma_u^+$, $1^2\Pi_u$, and $2^2\Pi_u$ dissociation channels. All these transitions are of the *continuum-continuum* type.

The above field-free analysis does not make it clear which states can contribute to the observed fragmentation processes. As stated in Refs. [1,10], a *dressed state* [18] picture allows for a more insightful understanding of the dynamics. Fig. 3 shows the *field-dressed* version of the potential curves. For a given electronic state $|\psi_i\rangle$ and a given photon number $|N-n\rangle$, designating an initial *N*-photon state from which *n* photons were removed by absorption, a *diabatic* field-dressed state can be constructed as $|\psi_i\rangle|N-n\rangle$ with a total energy of $V_i(R)+(N-n)\hbar\omega$. The initial state is represented as $|1^2\Sigma_g^+\rangle|N\rangle$ (i.e., $n = 0$) with the total energy of $V_{1^2\Sigma_g^+}(R)+N\hbar\omega$. Taking into account the fact that *ungerade* states can only be reached by an odd number of photons ($n$ = 1 or 3), while *gerade* states can only be reached by an even number of photons ($n$ = 0 or 2), our description incorporates 19 field-dressed states. In this way we account for all the possible transitions shown in Fig. 3.

Care must be taken in the curve-crossing regions (indicated by the circles in Fig. 3) occurring along the initial $|1^2\Sigma_g^+\rangle|N\rangle$ potential curve. These crossings become *avoided crossings* in the *adiabatic* representation [19]. Two $\Sigma$ type field-dressed curves cross the ground state potential at regions *c* (~7.2 Å) and *d* (~9.8 Å), which are at much larger internuclear distances than the excursion made by the dissociating wave packet during the laser pulse, whose 150 fs duration is equal to one-half the vibrational period of $Na_2^+$. Hence, these two states are unlikely to play a major role in the fragmentation processes. At region *b* (~5.4 – 5.9 Å), three other field-dressed curves cross the ground state potential. The two $\Pi$ states can be ruled out, because the $\Pi-\Sigma$ type transition cannot account for the observed fragment distribution which has to correlate with a $\Sigma-\Sigma$ type transition. The $\Sigma$ state crossing at region *b* can also be ruled out, mainly because it is coupled to the ground state via a third-order interaction only, and the crossing also occurs at relatively large distances.

The only possibility left is the three diabatic field-dressed states which cross the ground state potential at region *a* (~4.5 Å), and are connected to the ground state by the following ladder of successive *first-order* couplings:

$\left|1^2\Sigma_g^+\right\rangle|N\rangle \xrightarrow{\hbar\omega} \left|1^2\Sigma_u^+\right\rangle|N-1\rangle \xrightarrow{\hbar\omega} \left|2^2\Sigma_g^+\right\rangle|N-2\rangle \xrightarrow{\hbar\omega} \left|2^2\Sigma_u^+\right\rangle|N-3\rangle$. Given that the initial wave packet is prepared directly from the $X^1\Sigma_g^+$ ground state of Na$_2$ via a *vertical* non-resonant four-photon ionization, the initial dissociating wave packet in the ground ionic state continuum is expected to start evolving from roughly the same internuclear distance as the equilibrium internuclear distance of the $X^1\Sigma_g^+$ ground state ($R_e \sim 3.1$ Å). Compared to the other three regions (*b*, *c*, and *d*), multicrossing region *a* is the closest one with respect to $R_e(X^1\Sigma_g^+)$, hence the most probable one to be reached by the dissociating wave packet during the laser pulse duration. The three dressed states crossing at this region may give rise to three dissociation channels with absorption of *n* = 1, 2, or 3 photons. By computing the energy difference between this region and each dissociation asymptote for the three field-dressed states, we obtain that the three total fragment energies are: 0.73 eV for $\left|1^2\Sigma_u^+\right\rangle|N-1\rangle$, 0.17 eV for $\left|2^2\Sigma_g^+\right\rangle|N-2\rangle$, and 1.74 eV for $\left|2^2\Sigma_u^+\right\rangle|N-3\rangle$, respectively.

By comparison of the observed fragment kinetic energies with these theoretically predicted energies, we can now readily interpret the origin of each observed fragment channel as follows: (1) Channel *B* (0.8 ± 0.2 eV) can be attributable to a *one-photon* transition between the $1^2\Sigma_g^+$ and $1^2\Sigma_u^+$ states, which leads asymptotically to Na(3s) + Na$^+$, (2) Channel *A* (0.2 ± 0.05 eV) originates from a *two-photon* transition between the $1^2\Sigma_g^+$ and $2^2\Sigma_g^+$ states, which leads to Na(3p) + Na$^+$, (3) Channel *C* (1.8 ± 0.3 eV) arises from a *three-photon* transition between the $1^2\Sigma_g^+$ and $2^2\Sigma_u^+$ states, which also leads to Na(3p) + Na$^+$. The fragments resulting from the channels *A* and *C* are the same, and the separation of the fragment kinetic energies for these two channels (~1.6 eV) equals, within the experimental errors, one quantum of photon energy ($\hbar\omega$ ~1.55 eV), which indicates that channel *C* can be assigned as being due to ATD.

Although the initial ground state continuum exceeds its dissociation limit by ~0.35 eV ($n = 0$ line in Fig. 2), fragments with corresponding kinetic energy release have not been observed. Therefore channel *B* cannot be viewed as an ATD channel, simply because one photon is required, rather than in excess, for the commencement of the fragments Na(3s) + Na$^+$.

To further confirm the above interpretation, we have constructed more diabatic field-dressed states with higher-lying Σ states and transitions with a higher number (e.g., $n = 4$ and 5) photons, and have also examined other possibilities of initial Na$_2^+$ preparation schemes (e.g., five other than four-photon ionization). None of these scenarios can result in a reasonable agreement with the experimental observation.

It is worth noting that no fragmentation channels associated with field ionization followed by Coulomb explosion were observed, since the fragments from such channels are expected to carry much higher kinetic energies [7]. This implies that under our experimental conditions the ATD channel dominates, which conforms to the theoretical predictions in Ref. [6].

The physical processes involved in the three observed fragmentation channels can be elucidated in the following way: When the laser pulse is turned on, the system wave packet which exists initially in the ground state continuum, starts moving along the *n* = 1, 2, and 3 *light-induced potentials* towards larger internuclear distances. Already within the laser pulse duration this wave packet reaches a distance in the vicinity of 4.5 Å where branching occurs. By the time the laser pulse is turned off, the wave packet populates the repulsive light-induced potentials leading to dissociation and/or above-threshold dissociation.

It is conceptually helpful to think of such strong-field photodissociation as *laser-induced predissociation* [19]. Given that the laser pulse duration is in the order of one-half a vibrational period, the wave packet does not have enough time to move farther to larger distances, thereby closing the dissociation channels associated with the long range multicrossings.

In conclusion, we have reported the first observation of above-threshold dissociation of a multielectronic molecule – $Na_2^+$ – in an intense laser field via field-dressed continuum-continuum transitions. The involved photodissociation dynamics is conveniently visualized and well justified using a field-dressed four-state model.

This research was made possible by financial support from the Canada Foundation for Innovation. Q. Z. would like to thank the University of Science and Technology of China (USTC) for support grants. The authors have enjoyed useful discussions with G. Gerber. Experimental assistance from M. Keil, C. Li, S. Zhdanovich, V. Milner, and Q. Hu is also gratefully acknowledged.

[1] A. Giusti-Suzor, F. H. Mies, L. F. DiMauro, E. Charron, and B. Yang, J. Phys. B **28**, 309 (1995), and references therein.

[2] S. Chelkowski, T. Zuo, O. Atabek, and A. D. Bandrauk, Phys. Rev. A **52**, 2977 (1995).

[3] P. Dietrich, M. Yu. Ivanov, F. A. Ilkov, and P. B. Corkum, Phys. Rev. Lett. **77**, 4150 (1996).

[4] T. D. G. Walsh, F. A. Ilkev, and S. L. Chin, J. Phys. B **30**, 2167 (1997).

[5] C. Wunderlich, H. Figger, and T. W. Hänsch, Phys. Rev. A **62**, 023401 (2000).

[6] M. Machholm and A. Suzor-Weiner, J. Chem. Phys. **105**, 971 (1996).

[7] A. Assion, T. Baumert, U. Weichmann, and G. Gerber, Phys. Rev. Lett. **86**, 5695 (2001).

[8] J. McKenna *et al.*, Phys. Rev. Lett. **100**, 133001 (2008).

[9] R. Torres *et al.*, Phys. Rev. Lett. **98**, 203007 (2007).

[10] A. Giusti-Suzor, X. He, O. Atabek, and F. H. Mies, Phys. Rev. Lett. **64**, 515 (1990).

[11] N. K. Rahman, J. Phys. (Paris), Colloq. **46**, C1-249, 1985.

[12] P. H. Bucksbaum, A. Zavriyev, H. G. Muller, and D. W. Schumacher, Phys. Rev. Lett. **64**, 1883 (1990); A. Zavriyev, P. H. Bucksbaum, H. G. Muller, and D. W. Schumacher, Phys. Rev. A **42**, 5500 (1990); B. Yang, M. Saeed, L. F. DiMauro, A. Zavriyev, and P. H. Bucksbaum, Phys. Rev. A **44**, R1458 (1991).

[13] S. Magnier, M. Persico, and N. Rahman, Chem. Phys. Lett. **279**, 361 (1997).

[14] K. Sändig, H. Figger, and T. W. Hänsch, Phys. Rev. Lett. **85**, 4876 (2000).

[15] C. Trump, H. Rottke, and W. Sandner, Phys. Rev. A **59**, 2858 (1999).

[16] E. Charron, A. Giusti-Suzor, and F. H. Mies, Phys. Rev. A **49**, R641 (1994).

[17] A. Henriet, J. Phys. B **18**, 3085 (1985).

[18] A. D. Bandrauk, E. Aubanel, and J. M. Gauthier, in *Molecules in Laser Fields*, edited by A. D. Bandrauk (Marcel Dekker, New York, 1994), Chap. 3.

[19] A. M. F. Lau and C. K. Rhodes, Phys. Rev. A **16**, 2392 (1977).

**FIGURE CAPTIONS**

FIG. 1. Time-of-flight spectrum of $Na^+$ fragments obtained from fragmentation of $Na_2^+$ ($1^2\Sigma_g^+$) in an intense ($3 \times 10^{12}$ W/cm$^2$) 800 nm, 150 fs laser field. The three fragmentation channels are labeled *A*, *B*, and *C*, with the corresponding measured total released kinetic energies indicated.

FIG. 2. (Color online) Field-free potential energy curves for the 12 lowest electronic states of $Na_2^+$. The $\Sigma$ states are represented by solid lines, the $\Pi$ states by broken lines. The initial energy of the ion is indicated by the lowest horizontal dashed-dotted line ($n = 0$). The three higher lines correspond to absorption of n = 1, 2, 3 photons. The vertical arrows indicate a photon energy of ~1.55 eV for $\lambda = 800$ nm.

FIG. 3. (Color online) Diabatic field-dressed potential energy curves constructed from the first 12 electronic states of $Na_2^+$. The *ungerade* states are dressed with n = 1, 3 photons, and the *gerade* states with n = 2 photons. Four multi-crossing regions are indicated by solid circles, labeled *a*, *b*, *c*, and *d*. The branching from region *a* is of particular interest in this work.

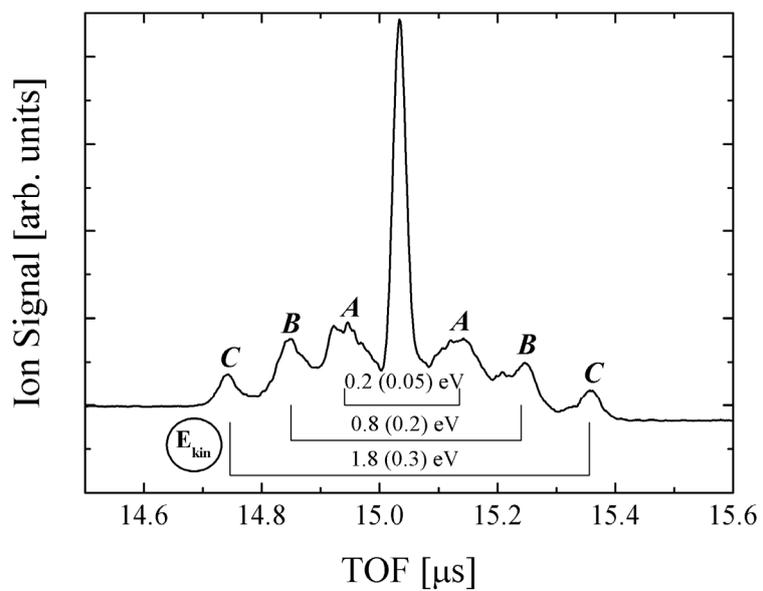

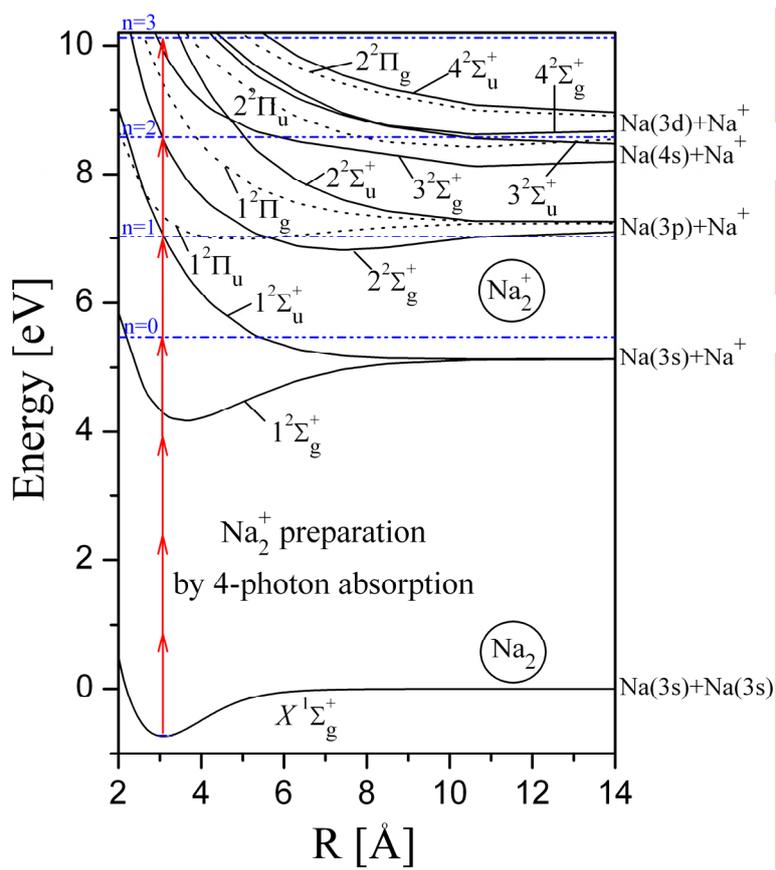

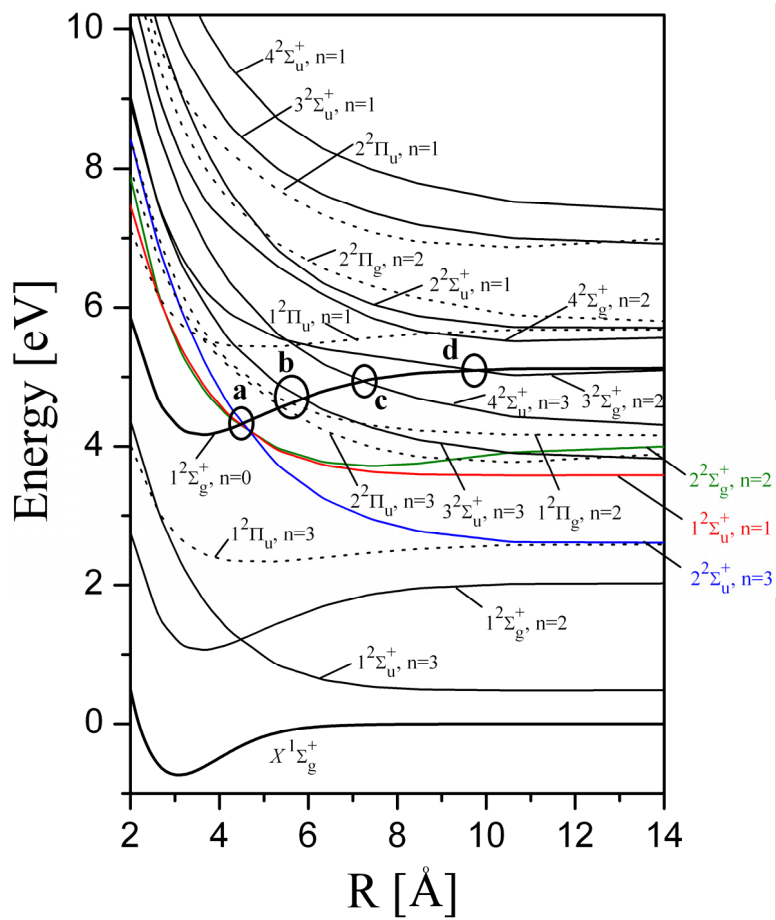